\documentstyle[editedvolume]{crckapb}

\input{epsf}

\begin{opening}
\title{STRONG GRAVITY AND X-RAY SPECTROSCOPY}

\author{A.\ MACIO{\L}EK-NIED\'ZWIECKI}
\institute{{\L}\'od\'z University, Department of Physics\\
           Pomorska 149/153, 90-236 {\L}\'od\'z, Poland
}
\author{P.\ MAGDZIARZ}
\institute{University of Durham, Department of Physics\\
          South Road, Durham DH1 3LE, UK
}
	
\end{opening}

\runningtitle{STRONG GRAVITY AND X-RAY SPECTROSCOPY}

\begin{document}

\begin{abstract}
This paper reviews the effects of general relativity in an X-ray spectrum
reflected from a cold matter accreting onto a black hole. The spectrum 
consists of the iron K$\alpha$ line and the Compton reflection. We sketch 
the overall picture of radiative processes in the central parts of the 
accretion flow with relation to the relativistic effects derived from the
discrete features in the X-ray spectrum. We discuss implications 
for detection of relativistic effects and computational tools of spectral 
analysis.
\end{abstract}

\section{Introduction}

Among various observational signatures of accreting black holes (e.g.,
Madejski 1998; Paradijs 1998) perhaps the most convincing source of
information on the strong gravitational field in the vicinity of the black
hole comes from X-ray spectroscopy. In the central part of the accreting
flow the emitting matter exists in at least two phases: a hot, mildly
relativistic, moderately optically thick ($\tau\sim 1$) plasma emitting
hard X-ray continuum (e.g., Poutanen 1998) and a cold, Compton-thick,
thermal plasma of temperatures characteristic for optical/UV/soft X-ray
energy range (e.g., Blaes 1998). Such two emitting phases of matter are
radiatively coupled to each other, leading to a complex variable geometry
of the accreting flow (e.g., \.Zycki, Done \& Smith 1998; Magdziarz et
al.\ 1998).  However, both the multi-temperature thermal continuum emitted
by the cold matter and the hard X-ray continuum produced by Comptonization
of the soft photons by the hot, thermal plasma are less useful for probing
the geometry of the emitting source: the first one, due to complex atomic
physics and superposition of emission from mostly more distant components,
and the second one due to basic featureless continuum character.  On the
other hand, illumination of the cold matter by the hard X-rays produces
Compton reflection component with superimposed discrete signatures coming
from bound-free absorption and fluorescence lines (e.g., George \& Fabian
1991).  The most pronounced of these are the iron absorption edge and the
iron K$\alpha$ fluorescence line, at 7.1 keV and 6.4 keV, respectively
(for the neutral matter), both discovered in the X-ray spectra of black
hole systems by {\it Ginga} (e.g., Pounds et al.\ 1990).  Modeling the
energy shifts of those spectral features (with respect to the known rest
energies) we can investigate physical conditions in the vicinity of the
black hole, where the generation of the X-ray luminosity takes place. {\it
ASCA} observations have confirmed that the X-ray spectra contain a
significant contribution from the matter emitting in the immediate
vicinity of the black hole, however we need a much higher quality data to
investigate the velocity field of that matter, its physical state, and the
gravitational potential in the emission region. 

In this paper we review the theoretical models developed for the analysis
of relativistic effects in the spectra generated by black hole accretion 
(section 2) and summarize the observational data indicating presence of such 
effects (section 3). We also discuss the possibility of the estimation of the
black hole spin by modelling the X-ray spectral features (section 4).

\section{Physical outline}

In order to investigate the relativistic distortion affecting the spectrum
of the radiation coming from the region with the strong gravitational
field, in particular to obtain the precise profiles of the discrete
features, one has to solve directly the geodesic equation. No analytical
solutions are available apart from the Fabian et al.\ (1989) formulae for
the emission line profile from a disk around a Schwarzschild black hole.
However, these formulae fail in the strong field limit. Therefore, the
calculation of the relativistic effects requires quite complex numerical
procedures. 

Since the publication of the paper by Bardeen (1970) it became clear that
black holes in accretion powered systems should be described by the Kerr
metric rather than the less general Schwarzschild metric, as the accretion
is likely to spin up the black hole. The Kerr black hole can be
characterized by two parameters, the mass, $M$, and the dimensionless spin
parameter, $a$.  The maximum value of $a=0.998$ cannot be exceeded in
accretion systems (Thorne 1974) due to a torque exerted on the hole by
radiation from the accreting matter (cf. Moderski \& Sikora 1996 for the
recent calculation of the black hole spin evolution). 

All the relativistic effects affecting the spectra, except the variability
time-scale, do not depend on the absolute distance from the black hole,
and are functions of the distance in units of the gravitational radius,
$r_{\rm g}=GM/c^2$, only.  Therefore, the same quantitative effects would
occur for both the stellar mass ($10 M_\odot$) and supermassive ($10^6-10^9
M_\odot$) black holes. The value of the spin influences the trajectories
close to the black hole yielding the distance of the last (quasi) stable
circular orbit $r_{\rm ms}= 6 r_{\rm g}$ for $a=0$ and $r_{\rm ms}=1.23
r_{\rm g}$ for $a=0.998$. 

The velocity field of the accreting matter is usually treated in the
approximated way in models of relativistic effects: 
it is assumed that the matter accretes in a form of a
geometrically thin accretion disk (cf.\ Jaroszy\'nski \& Kurpiewski 1997
for the observational effects in geometrically thick disks expected in
low-efficiency systems) and, for $a>0$, it is assumed that the disk
rotates in the planar plane of the black hole (cf.\ Bardeen \& Petterson
1975 for frame dragging effects on disks not in the equatorial plane). The
accreting material is assumed to flow along circular equatorial geodetics
with a superimposed small radial inflow (neglected in calculations) for
$r>r_{\rm ms}$, and to be in free fall for $r<r_{\rm ms}$. 
 
The first calculation of the spectrum of the matter accreting onto a Kerr
black hole was presented by Cunningham (1975). His results, however, do
not have the form suitable for data analysis, as they are tabulated only
for a few values of the relevant parameters. So far, the most efficient
procedure for the data analysis comes from the concept of the photon
transfer function, which is constructed by calculating the trajectories of
a large number of photons (e.g., Laor 1991). 
  
 A further issue is effect of the relativistic transfer within the source.
As pointed out by Cunningham (1976) the returning radiation may give an
additional contribution to the illumination of the accretion disk. This
effect was studied by Dabrowski et al.\ (1997) who found that the
returning disk radiation in the Kerr metric gives no significant effect,
in particular that it cannot enhance the equivalent width of the line by
more than 20 \%.  On the other hand, the light bending may become very
important when one considers the transfer of the radiation from the X-ray
source to the disk, leading to strong enhancement of the equivalent width
of the iron line (Martocchia \& Matt 1996) and the amount of the Compton
reflected radiation. 

\begin{figure} 
\begin{center} \leavevmode \epsfysize=6.5truecm\epsfbox{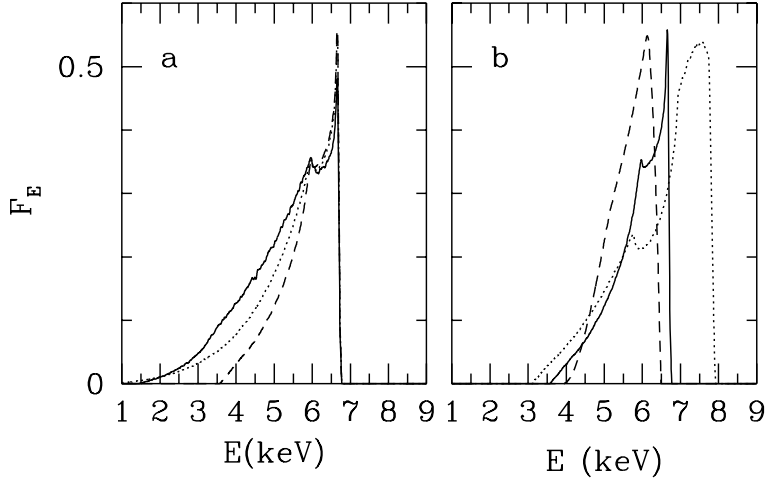}
\end{center}
\caption{ a) The dependence of the line profile on the geometry of the
innermost part of the accretion flow. The rest energy of the line is
assumed to be 6.4 keV (neutral iron K$\alpha$ line). All the curves give
the line profiles for the inclination angle 30$^{\rm o}$, the outer radius
of the emission region, $r_{\rm out}=100 r_{\rm g}$, and the radial
emission law favoring the emission from the central parts (the local
intensity of the emission of the line $I(r) \sim r^{-3}$). The dashed and
dotted curves correspond to the Schwarzschild geometry with the inner
radii, $r_{\rm in}=6 r_{\rm g}$ and 2$r_{\rm g}$, respectively (the latter
includes the contribution from the free-falling material). The solid curve
corresponds to the Kerr geometry ($a=0.998$) with $r_{\rm in}=1.3r_{\rm
g}$. The Kerr metric disk gives more redshifted photons due to the
location of the inner edge of the disk. Then, the high-$a$ line has an
excess in the red tail, with respect to the low-$a$ line, if only the
fluorescence from the circularly flowing disk material is taken into
account.  However, the contribution from the free-falling matter in the
Schwarzschild metric removes this robust difference between the line
profiles.  b) The dependence of the line profile on the inclination angle.
The curves correspond to the disk in the Schwarzshild metric extending
between 6$r_{\rm g}$ and 100$r_{\rm g}$ with $I(r) \sim r^{-3}$, observed
at 60$^{\rm o}$ (dotted curve), 30$^{\rm o}$ (solid curve) and 10$^{\rm
o}$ (dashed curve).  The position of the high energy peak of the line and
the extent of the red tail can be used for the precise determination of
the inclination and the distance of the emission region.}
\label{lines} 
\end{figure}

\begin{figure}
\begin{center}
\leavevmode
\epsfysize=6.5truecm\epsfbox{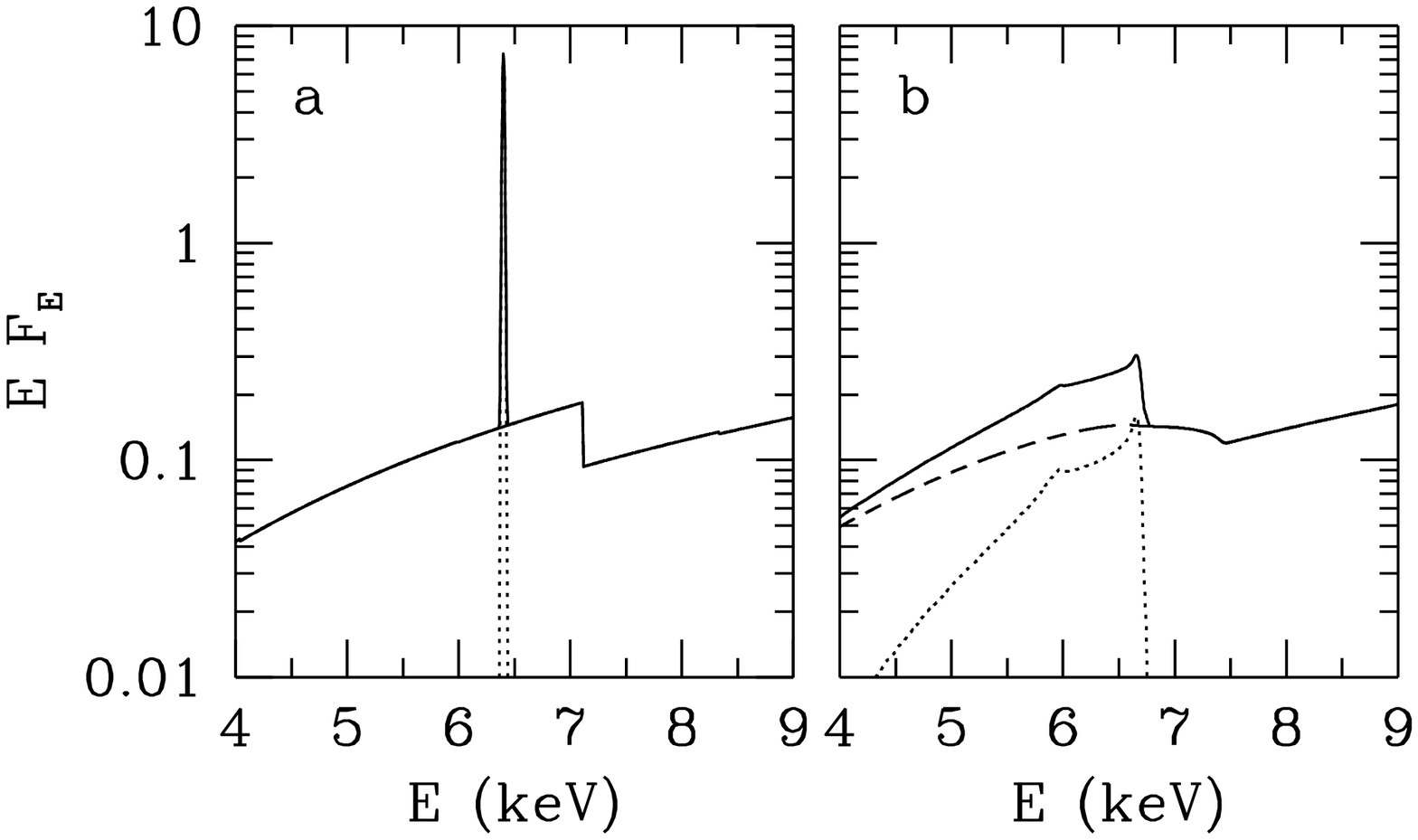}
\end{center}
\caption{The figure compares the shape of the iron K$\alpha$ line and
absorption edge in the reflection spectrum arising from the X-ray
irradiation of the slab at rest (a) with those in the reflection from the
innermost region of the accretion disk in the Schwarzschild geometry (b)
with $r_{\rm in}=6 r_{\rm g}$, $r_{\rm out}=100 r_{\rm g}$, $I(r) \sim
r^{-3}$ and $\cos \theta_{\rm obs}=30^{\rm o}$. The solid curve shows the
sum of the line (dotted curve; $EW=150$ eV in both geometries) and the
Compton reflection (dashed curve). The relativistic transfer effects smear
both spectral components, which add up giving rise to the complex spectral
feature around 7 keV.}
\label{reflect}
\end{figure}

Figure 1 shows the emission line profiles from the accretion flows. The
lines are typically broad and skewed rather than the double-peaked
structures derived in the Newtonian approximation (cf.\ Fabian 1997 for
the basic discussion of the effects in the line profiles). In general the
characteristics of the lines allow to distinguish them from the line
shapes produced by other physical processes (Fabian et al.\ 1995). The
precisely measured line profile can be used for the determination of such
system parameters as the inclination or the location of the emission
region.  However, the analysis of relativistic effects may likely lead to
unreliable results, since modelling of the iron line and the iron edge is
strongly dependent of the continuum model in most X-ray data (e.g.,
Zdziarski, Johnson \& Magdziarz 1996). X-ray spectral deconvolution is a
non-linear, complex procedure and it cannot be simplified by any indirect
method of the analysis of the line profile. E.g., it is possible to
constrain the parameters of the system using the extreme frequency shifts
of the line profile (Bromley, Miller \& Pariev 1998), however,
determination of these extreme shifts requires prior deconvolution of the
line profile from the data, which yields by itself the best fitting set of
the parameters of a given assumed model. 
 
A particularly important, but usually neglected, effect is due to the
relativistic distortion of the Compton reflected radiation (cf.\ Figure
2). The reflected continuum is affected similarly to the fluorescent line
since it originates from the same matter (cf.\ Matt, Perola \& Piro 1991),
although the dependence of reflection and fluorescence on ionization is
different, and they are not necessarily smeared in the same manner. A
further shortcoming of the usually applied models is the simplified
treatment of the fluorescent line, which is approximated as a
$\delta$-function.  Due to Compton scattering of fluorescent photons in
the disk, the line flux emerges locally as a broadened spectral feature,
which effect is especially pronounced in highly ionized disks (e.g., Matt,
Fabian \& Ross 1996). The above problems will be addressed in detail in
our forthcoming paper (Macio\l ek-Nied\'zwiecki \& Magdziarz 1998). The
spectral components, relevant to the determination of the iron line
profile, are shown in the best fit of our model to the {\it Ginga} data of
the Seyfert 1 galaxy MCG-6-30-15. 

\begin{figure} 
\begin{center} \leavevmode \epsfysize=5.5truecm\epsfbox{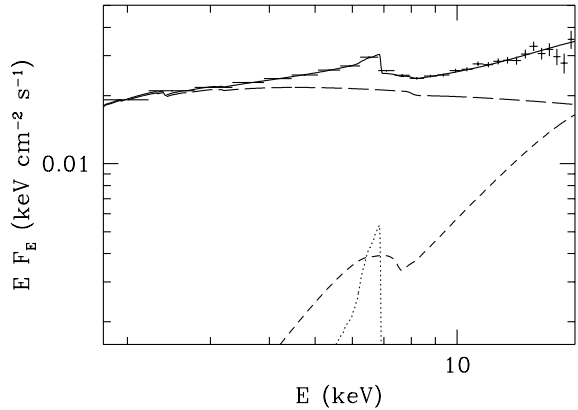}
\end{center}
\caption{The average {\it Ginga} spectrum of MCG-6-30-15.  The
fitted model (solid curve) is a power-law (the energy spectral index
$\alpha=1.1$), attenuated by the ionized absorber (long-dashed curve), the
Compton reflection with the amplitude corresponding to the solid angle of
the reprocessor as seen from the X-ray source of $3 \pi$ (for isotropic
X-ray illumination; short dashed curve) and the Fe K$\alpha$ line with
$EW=300$eV (dotted curve). All the components are self-consistently
transferred through the Schwarzschild metric field with the assumption
that the emission comes from the region between 6$r_{\rm g}$ and
100$r_{\rm g}$. The fitted radial emission law, $I(r) \sim r^{-3.9}$, and
the inclination $\theta_{\rm obs}=40^{\rm o}$. The fit weakly prefers the
relativistically smeared spectrum over a reflection from a distant
reprocessor. Furthermore, the distant reprocessor requires nearly edge-on
inclination (in contradiction with its Seyfert 1 classification), 
as well as the
solid angle subtended by the reflector $\sim$$10\pi$, which would require
a complex model with strongly obscured X-ray source. On the other hand, the
parameters of the relativistic model would be compatible with a nearly
flat disk, provided that the isotropic X-ray source is close to the black hole
horizon. In this case the light bending effects can account for both the
very steep illumination of the disk (as implied by the fitted 
radial emissivity) and the enhanced amplitude of the iron line and Compton
reflection.}
\label{ginga}
\end{figure}

Finally, the innermost region of the accretion flow appears to be
extremaly complex. It is well known, that solutions of the standard
accretion disk with dissipation rate proportional to pressure are unstable
(Shakura \& Sunyaev 1976) likely leading to the disk fragmentation (e.g.,
Krolik 1998). Such a complex structure of the central region has been
indeed invoked to explain both the nature of EUV emission in AGNs in
general (e.g., Kuncic, Celotti \& Rees 1997) and the overall spectral
phenomenology in some particular sources (e.g., Magdziarz \& Blaes 1998;
for NGC~5548). The complex radiative coupling of the multi-phase medium
convolved with relativistic dynamics seems to be plausible alternative to
a simple disk geometry. However, the profiles of the discrete features in
such geometry need more complex treatment, and have not been calculated as
yet. 

\section{Observational outline} 

The presence of the broad iron lines, with profiles indicating strong
relativistic effects, has been found in a number of Seyfert 1s observed by
{\it ASCA} (e.g., Nandra et al.\ 1997a). Such objects show no spectral
contamination from jet activity and have relatively low absorption in the
direction of the central part of the accretion flow. The iron line has
typically $EW \sim $200 eV and the profile consistent with that expected
from an accretion disk with a bulk of emission coming from a region within
$\sim$20$r_{\rm g}$. The observed peak energy at 6.4 keV suggests that the
emission comes mainly from nearly neutral matter, although the observed
anticorrelation of the equivalent width of the broad line with the source
luminosity (Nandra et al.\ 1997b) indicates that the reflecting matter
tends to be more ionized in more luminous sources. It is also likely that
the profiles contain a small (EW$\sim$25 eV) constant and narrow component
from a more distant reprocessor, e.g., a molecular torus. Orientation of
the disk surface inferred for the sample of Seyfert 1 galaxies seems to be
consistent with the unification scheme (Antonucci \& Miller 1985),
however, an analysis of a sample of Seyfert 2s suggests that either a
population of the obscured objects or geometry of the central part of the
accretion flow is more complex (Turner et al.\ 1998; Elvis et al.\ 1998).
We note, however, that the above results are likely to be subject to the
modelling problems indicated in section 2. 

The most convincing example of relativistic broadening of the Fe K$\alpha$
line comes from observations of the Seyfert 1 galaxy MCG-6-30-15 (Tanaka
et al.\ 1995; cf.\ Figure 4). The profile seems to vary with the source
luminosity in the sense that it shows stronger relativistic effects in
fainter states of the source (Iwasawa et al.\ 1996). At the minimum state
the profile seems to require either models with fluorescence from matter
free-falling below the last stable orbit or models with maximally rotating
black hole (cf.\ section 4). The signatures of the relativistic smearing
of both the iron line and absorption edge have been also found in {\it
Ginga} observations of Galactic black hole binaries: V404 Cyg (\.Zycki,
Done \& Smith 1997) and Nova Muscae (\.Zycki, Done \& Smith 1998). 

\begin{figure}
\begin{center}
\leavevmode
\epsfysize=4.6truecm\epsfbox{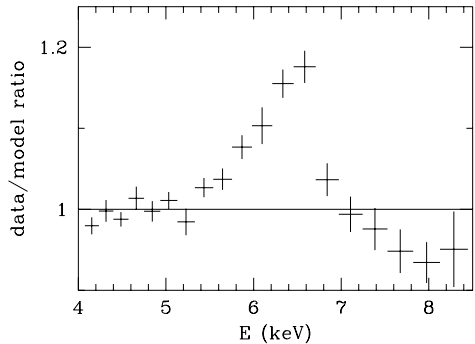}
\end{center}
\caption{A ratio of the average data over power-law model for the {\it
ASCA} observation of MCG-6-30-15 (July 1994). All four {\it ASCA} detectors
where taken into account to achieve high signal-to-noise ratio which is
sufficient to reveal the skewed profile of the line, with the maximum at
6.6 keV and the broad tail at lower energies, independently of the assumed
model of the continuum. The iron absorption edge above 7 keV appears
consistent with relativistic smearing.}
\label{asca}
\end{figure}

\section{The value of the black hole spin}

The spin of the black hole appears to be crucial for the properties of
accretion powered objects, according to the unification scheme of AGNs, in
which one expects that radio-loud jet dominated active galaxies host
rapidly rotating black holes (cf.\ Blandford 1990). To test this
unification model, one needs an independent measurement of the spin value.
The mass of the black hole can be directly determined from the orbital
motion in some X-ray binaries. A lower limit on $M$ can be derived for
AGNs from the requirement that their luminosity cannot exceed the
corresponding Eddington limit.  On the other hand, the estimation of the
spin is based on more sophisticated arguments. The spectrum of the
accretion disk depends on $a$ (Novikov \& Thorne 1973), however, we do not
understand the reprocessing of the disk radiation sufficiently well to be
able to constrain the spin value from the shape of the accretion disk
spectrum, although the shape of the ultrasoft component in Galactic
sources with superluminal jets may indicate that they are powered by
rapidly rotating black holes (Zhang, Cui \& Chen 1997). 

The effect of the black hole rotation on the line profile has been
investigated, e.g., by Kojima (1991) who did not find any significant
dependence of the line profile on $a$ if fluorescence occurs at a distance
$r > 6 r_{\rm g}$. As for the rapidly rotating black hole the disk extends
well below $6r_{\rm g}$, the emission from that region gives the extra
flux in the red wing of the line, which makes the low-$a$ and high-$a$
lines clearly distinguishable (see Figure 1). Reynolds \& Begelman (1997)
have pointed out, however, that fluorescence by the material inside the
radius of marginal stability can have an observable influence on the iron
line profile causing the red wing to be much wider.  In fact,
Jaroszy\'nski (1997) finds that the difference in the velocity field, in
the innermost region, between low-$a$ and high-$a$ cases does not yield
any robust difference in the line profiles. Given the uncertainty in such
parameters of the system as the radial emissivity law or inclination
(which is not independently constrained, except for some black hole
binaries, and must be found in the fitting procedure) we cannot expect to
constrain the value of $a$ from the line profile with the present or even
forthcoming X-ray data quality, unless we better understand the properties
and geometry of the matter in the region inside $6 r_{\rm g}$. In fact,
the model by Reynolds \& Begelman (1997) implies that the free-falling
matter gets completely ionized and cannot contribute to the line profile,
except for very inefficient X-ray sources. The efficiency as low as $\eta
\sim 10^{-5}$ is needed (where $L_{\rm X} \sim \eta \dot m c^2 $, $L_{\rm
X}$ is the X-ray luminosity and $\dot m$ is the accretion rate) to avoid
the complete ionization of that region. Such a low efficiency is very
unlikely in the luminous accretion powered systems.  Furthermore, the
model by Reynolds \& Begelman (1997) relies on the central location of the
X-ray source, which could be the case if the X-ray emission was related to
the formation of the jet activity. However, if the emission is connected
locally with the dissipation of the energy in the disk (as in the model
with magnetic flares above the disk surface), one may expect that the line
will be dominated by the flourescence occurring in the disk. There is
virtually no dissipation of energy once the material has passed the last
stable orbit, so in this case the contribution from the free-falling
matter should not be very significant, independently of its ionization
stage. 

If any of the above arguments allows us to exclude the contribution from
the free-falling matter, the observation of the very extended red tail of
the line can be used as an unambiguous evidence for the rapid rotation of
the central black hole. 

\section{Summary} 

X-ray spectroscopy offers, for the first time, the unique opportunity to
investigate the geometry of the region in the immediate vicinity of the
black hole horizon. Currently available data suggest that the observed
X-ray spectra contain a significant contribution from matter emitting in
the vicinity of the black hole, however, results are still uncertain
mostly due to complex response of X-ray instruments, which confuses the
iron line profile with both the iron edge and the underlying continuum. 
The progress in determination of the iron line and, consequently, the
innermost geometry of the accretion flow, is expected to result from new
X-ray observatories of high spectral resolution, to be launched in the
near future. 

\medskip\noindent

{\it Acknowledgements}\/ We are grateful to Greg Madejski for helpful
discussion and his hospitality during our stay in NASA/GSFC where this
paper has been written. We are also grateful to Chris Done for her
comments and reviewing the manuscript.

\end{document}